\documentclass[twocolumn, aps, amssymb, amsfonts]{revtex4}

\usepackage{epsfig}
\newcommand{\tr}{{\rm tr}}

\begin{document}

\title{
Nonergodicity of entanglement and its complementary behavior to magnetization in infinite spin chain} 

\author{Aditi Sen(De),  Ujjwal Sen, and Maciej Lewenstein}
\affiliation{Institut f{\" u}r Theoretische Physik, Universit{\" a}t Hannover, 
D-30167 Hannover, Germany}

\begin{abstract}

We consider the problem of the validity of a statistical mechanical description of 
two-site entanglement in an infinite spin chain described by the XY model Hamiltonian. We show that 
the two-site entanglement of the state, evolved from the initial equilibrium state, after a 
change of the magnetic field, does 
not approach its equilibrium value. This suggests that two-site entanglement, like (single-site) magnetization, 
is a nonergodic quantity in this model. Moreover we show that these two nonergodic quantities 
behave in a complementary way.  

\end{abstract}

\maketitle

\def\com#1{{\tt [\hskip.5cm #1 \hskip.5cm ]}}

Entanglement plays a key role in the rapidly growing field of quantum information processing \cite{NC}, where
so far, most of the discussions of entanglement concern few body systems. 
%
%
%
%
%
%
%
Recently, however, the properties of entanglement has been also used to study and 
understand behavior of ``complex'' quantum systems. 
For instance, entanglement in quantum many body systems such as spin chains 
or Bose-Einstein condensates
were studied (see e.g. \cite{Osterloh, Nielsen, BEC}).
The role of ``entanglement length'' in quantum phase transitions 
was pointed out in Ref. \cite{ent_length}.

The aim of this paper is to deal with the statistical  properties of 
entanglement, or rather the validity of a statistical mechanical approach to entanglement 
in  realistic many body systems.
The relation of entanglement with important notions in statistical mechanics is important to 
understand the behavior of entanglement in quantum many body systems. 
We will restrict ourselves to one-dimensional infinite spin systems. 
Entanglement in such systems was considered e.g. in  Refs. \cite{Osterloh, Nielsen, Wootters}.
These papers deal exclusively with  the properties
of entanglement of an infinite spin chain, which is either 
in a ground state or in a thermal equilibrium state. Such states are ``static'' states of the system, as they 
do not explicitly depend on time.  Exceptions include the recent studies of 
entanglement in quantum dynamics, which allowed to formulate novel types of unprecedentedly
efficient numerical codes for simulation of quantum systems \cite{numerical} 
(see also \cite{talk_dilo_Osterloh, Briegel_orey_orey}).

The validity of a statistical mechanical 
description
of a 
quantity,
characterizing
a physical system, depends on the behavior of that quantity
as the system evolves in time. 
More precisely, a necessary condition for validity of  statistical mechanical description 
of a physical quantity is that it must be ``ergodic''. A physical quantity 
is said to be ergodic if the time average of the quantity matches its 
ensemble average. This is usually a hard question to check, even for classical 
systems. However, an indication of  whether a given quantity is  
ergodic or not, can be obtained by comparing the time evolved state with the 
equilibrium state.

To deal with such questions, we will therefore consider 
real
time evolution of 
an infinite one-dimensional spin chain. We will suppose that 
the spin chain is described by the so-called XY model (see 
Eq. (\ref{Hamiltonian_f_big})
below). We will show that for large times, the 
nearest neighbor entanglement of the evolved states,  does not 
approach to the corresponding entanglement of the equilibrium state. This suggests
that  nearest neighbor entanglement in this spin system is \emph{nonergodic} \cite{next-nearest},
indicating that one cannot describe entanglement in such models by equilibrium statistical mechanics.
Temporal dynamics of entanglement in spin systems is important in quantum information and computation tasks. 
For instance, the temporal dynamics of entanglement in a spin system has been used in Ref. \cite{Briegel_orey_orey} 
to obtain the 
one-way quantum computer. 
For the infinite  spin chain in the XY model, it is known that the single-site magnetization is 
also a nonergodic observable \cite{McCoy1}.
We will  show that (two-site) entanglement has 
a complementary temporal behavior to (single-site) magnetization in this model.
We will also show
that the 
(two-site) entanglement and magnetization of the evolved state 
saturates
for low temperatures,
for a given time.

The XY model (see Eq. (\ref{Hamiltonian_f_big})) that we consider here is integrable \cite{LSM}. 
For 
classical systems, integrable models possess
a 
large number of constants of motion, and are 
 usually not ergodic (see e.g. \cite{Gutzwiller}). 
However, such observations are valid for the case of systems that have a finite
 (and usually small) number of degrees 
of freedom. In our case, we deal with an infinite spin chain,
and
our 
considerations are distinctly quantum. Hence, questions about ergodicity, or its absence are nontrivial.

A one dimensional spin system (a one-dimensional array (lattice) of
spin-1/2 particles) 
with nearest neighbor interactions is described by a (dimensionless) Hamiltonian of the form
\(H_{int} = \sum_{i} ({\cal A} S_i^x S_{i+1}^x + {\cal B} S_i^y S_{i+1}^y + {\cal C} S_i^z S_{i+1}^z)\),
where the \(S_i^x\), \(S_i^y\), \(S_i^z\) are one-half of the Pauli spin matrices 
\(\sigma^x\), \(\sigma^y\), \(\sigma^z\) at the \(i\)-th site of the array, 
and \({\cal A}, {\cal B}, {\cal C}\) are 
coupling constants. Here, we take \({\cal A} \ne {\cal B}\) and \( {\cal C}=0 \).
We introduce also an external field into the Hamiltonian, so that the 
total Hamiltonian of the system is \(H(t) = H_{int} - h(t)H_{mag}\). To ensure 
that this external field has nontrivial effects on the evolution, we must have 
\( [H_{int}, H_{mag}] \ne 0 \). The simplest way in which this can be effected is 
by choosing \(H_{mag} = \sum_i S_i^z\), and \({\cal A} = 1+ \gamma\), \({\cal B} = 1 - \gamma\), 
\(\gamma \ne 0\).  
In this way we arrive at the XY model in the transverse field
%
%
\begin{equation}
\label{Hamiltonian_f_big}
H(t) = \sum_{i} \left[ (1 + \gamma) S_i^x S_{i+1}^x +  (1- \gamma) S_i^y S_{i+1}^y -  h(t) S_i^z \right],
\end{equation} 
where \(\gamma \ne 0\). In the following, we set \(\hbar =1\).
A spin chain whose dynamics is described by \(H(t)\), is said to be described by the ``XY model''.
For a finite number, \(N\), of spins, we assume 
a periodic boundary condition, \(\vec{S}_{N+1} = \vec{S}_1\). 
Ultimately, we will be interested in the thermodynamic limit \(N \rightarrow \infty\).
Such systems can be realized in atomic gas in an optical lattice (see e.g. 
\cite{proposal_korechhey}).


At a given time \(t\), the 
(thermal) equilibrium state is given by \(\rho_\beta^{eq}(t) = \exp (- \beta H(t))/Z\),
where \(Z=\tr(\exp (- \beta H(t)))\).
Here \(\beta = 1/kT\), where 
\(k\) is the Boltzmann constant, and \(T\) denotes the (absolute) temperature. 
To consider 
questions about ergodicity, 
we will be interested in the behavior of the evolved state. The evolution
is governed by the Hamiltonian \(H(t)\), from a given 
initial state.
Since we will compare the
properties of the evolved state, with those of the equilibrium state, it is natural to suppose that 
the initial state (at \(t=0\)),  is the equilibrium state at  \(t=0\). We denote the evolved state by
\(\rho_\alpha(t)\), where the suffix corresponds to the temperature of the initial equilibrium state 
\(\rho_\alpha^{eq}(0) = \rho_\alpha(0)\). 
For later times, properties of the evolved state \(\rho_\alpha(t)\) will be compared to 
those of the equilibrium state \(\rho_\beta^{eq}(t)\), so that they have the same energies: 
\begin{equation}
\label{energy}
\mbox{tr} (H(t) \rho_\alpha(t)) = \mbox{tr} (H(t) \rho_\beta^{eq}(t)).
\end{equation}
For simplicity, we 
will consider the case of sudden switch of the field \(h(t)\), as
\(
 h(t) = a\), for \(t \leq 0\),  
     and \( = b\), for \( t>0\).

Both \(\rho_\alpha(t)\) and \(\rho_\beta^{eq}(t)\) are states of an infinite number of spin-1/2 particles. 
For our purposes, it will be sufficient to consider single-site and two-site density 
matrices. Let us first consider the single-site density matrix for the state \(\rho_\beta^{eq}(t)\). 
By symmetry, the single-site density matrices of the chain are all the same. 
We will denote it by \(\rho^{eq}_1(t)\) (hiding the suffix \(\beta\)).
Now \(\rho^{eq*}_1(t) = \rho^{eq}_1(t)\), when the complex conjugation 
is taken in the computational basis, which (for each site) is the eigenbasis of Pauli matrix \(\sigma_z\).
Therefore \(\tr (S^y \rho^{eq}_1(t)) =0\). Moreover the Hamiltonian \(H(t)\) has the global 
phase flip symmetry (\([H,\Pi_i S^z_i]=0\)), from which it follows that \(\tr (S^x \rho^{eq}_1(t)) =0\). 
Consequently, the single-site density matrix of the equilibrium state is of the form 
\(
\rho^{eq}_1(t) = \frac{1}{2}I + 2 M^{eq}_z(t)S^z 
\),
where \(I\) is the \(2 \times 2\) identity matrix.
The evolved state does not necessarily have the property of being equal to its complex conjugation,
and consideration of the global phase flip symmetry is complicated 
by the fact that the Hamiltonian is explicitly dependent on time.
%
However, using the Wick's theorem, as in \cite{LSM,McCoy1,McCoy_eka},
the single-site density of the evolved state turns out to be of the form
\(
\rho_1(t) = \frac{1}{2}I + 2 M_z(t)S^z 
\).

For the case of the two-site density matrix 
\(\rho^{eq}_{12}(t)\) of the 
equilibrium state \(\rho_\beta^{eq}(t)\), we can again use the global phase flip symmetry and 
the fact that it is equal to its complex conjugate, so that it is of the form  
\begin{eqnarray}
\label{eq_two_site_state}
\rho^{eq}_{12}(t) = \frac{1}{4} I\otimes I + 
M^{eq}_z(t) (S^z \otimes I + I \otimes S^z)  \nonumber \\
 + \sum_{j=x,y,z} T^{eq}_{jj}(t) 
S^j \otimes S^j ,  
\end{eqnarray}
where 
the correlation functions, \(T^{eq}_{jj}(t)\), are  defined as
\(
T^{eq}_{jj}(t)= 4 \tr(S^{j}\otimes  S^{j} \rho^{eq}_{12}(t)), \quad j = x,y,z
\).
In the case of the two-site density matrix of the evolved state,
the \(yz\) and \(zx\) correlations are absent (via use of the Wick's theorem). 
However the \(xy\) correlations does not vanish, just as the \(xx\), \(yy\), and \(zz\) correlations.
Thus the two-site density matrix of the evolved state \(\rho_\alpha(t)\) is of the form
\begin{eqnarray}
\label{evol_two_site_state}
 \rho_{12}(t) = \frac{1}{4}  I\otimes I + 
M_z(t) (S^z \otimes I + I \otimes S^z)  \nonumber \\
+ T_{xy}(t) (S^x \otimes S^y + S^y \otimes S^x )
+ \sum_{j=x,y,z} T_{jj}(t) 
S^j \otimes S^j ,  
\end{eqnarray}
where 
the correlation functions, \(T_{jk}(t)\), are  defined as
\(
T_{jk}(t)= 4 \tr(S^{j}\otimes  S^{k} \rho_{12}(t)), \quad j,k = x,y,z
\).

Let us
consider 
the (single-site) 
magnetization  and the (two-site) entanglement in an infinite chain for the two states of interest. 
The magnetization of the equilibrium state is \(M^{eq}_z(t) = \tr(S^z \rho^{eq}_1(t))\),
while that for the evolved state is \(M_z(t) = \tr(S^z \rho_1(t))\). 
%
For studying entanglement \cite{footnote123, MichalQIC}, 
%
%
%
%
we will use logarithmic negativity (LN) \cite{VidalWerner} as our measure of entanglement.
LN of a bipartite state \(\rho_{AB}\) is defined as 
\(E_N(\rho_{AB})= \log_2 \|\rho_{AB}^{T_{A}}\|_1\),
where \(\|.\|_1\) is the trace norm, and \(\rho_{AB}^{T_{A}}\) denotes the partial transpose of 
\(\rho_{AB}\) with respect to the \(A\)-part \cite{Peres_Horodecki}. 
Note that the two-site density matrices in our case acts on \(\mathbb{C}^2 \otimes \mathbb{C}^2\).
Consequently, a positive value of the LN implies that the state is 
entangled and distillable \cite{Peres_Horodecki, Horodecki_distillable}, while 
\(E_N =0\) implies that the state is separable \cite{Peres_Horodecki}.


The (single-site)
magnetizations of the equilibrium state and the evolved state are respectively given by 
\cite{McCoy1}
\begin{equation}
\label{mag_initial}
M^{eq}_z(t) = \frac{1}{2 \pi} \int_{0}^{\pi}  d \phi \frac{\mbox{tanh}\left(\frac{1}{2} 
\beta \Lambda(h(t))\right)}{\Lambda(h(t))} (h(t) - \cos\phi), 
\end{equation}
and 
\begin{eqnarray}
\label{eq_magnetization}
 M_z(t) &=&  \frac{1}{2 \pi} \int_{0}^{\pi} d\phi
\frac{\mbox{tanh}(\frac{1}{2} \beta \Lambda(a))}{\Lambda(a) \Lambda^2 (b)} \nonumber \\
&\times &  \big[\cos(2 \Lambda(b) t) \gamma^2 (a-b) \sin^2\phi \nonumber \\
&-& (\cos \phi -b)[(\cos\phi-a)(\cos\phi-b) + \gamma^2 \sin^2 \phi \big],   \nonumber \\ 
\end{eqnarray}
where \(\Lambda(a)\) and \(\Lambda(b)\) 
can be obtained from 
\(
\Lambda(h(t)) = [\gamma^2 \sin^2 \phi + (h(t) - \cos \phi)^2]^{\frac{1}{2}}
\).

\begin{figure}[tbp]
\begin{center}
\epsfig{figure= 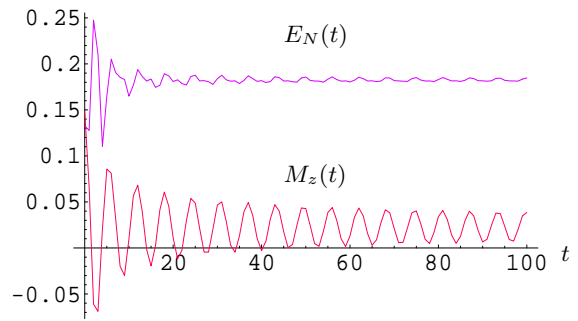,
height=.2\textheight,width=0.4\textwidth}
\put(-100,113){\(E_N(t)\)}
\put(5,30){\(t\)}
\put(-100,60){\(M_z(t)\)}
\caption{
Comparison of temporal
behavior of the nearest neighbor entanglement to the  magnetization, with time, 
of the evolved state of the  infinite spin chain in XY model:
We plot \(E_N(t)\) and \(M_z(t)\) for 
\(\gamma = 0.5\), \(\alpha = 200\), \( a=0.5\), \( b=0\). 
}
\label{M_ent_XY}
 \end{center}
\end{figure}

For definiteness, let us consider the case where 
\(
\gamma = 0.5\), \( \alpha = 200\), \( a=0.5\), \( b=0\). 
For these values of the parameters, Eq. (\ref{energy}) gives \(\beta \approx 3.9\).
 The equilibrium magnetization \(M_z^{eq}(t)\) 
has an initial value of \(\approx 0.148328\) and then jumps \emph{down} to zero for all later times. 
The magnetization \(M_z(t)\) of the evolved state, on the other hand, is an oscillating function
(see Fig. \ref{M_ent_XY}), which of course starts at the same initial value \cite{hotei_hobey} 
as \(M_z^{eq}(t)\), but remains positive
for long times. This suggests that the magnetization of the spin system in the XY model is 
 nonergodic \cite{McCoy1} (cf. \cite{Mazur}).

The nearest neighbor correlations for the equilibrium state 
are given by 
\cite{
McCoy_eka,two-site}
\(T^{eq}_{xx}(t) = - G^{eq}(-1,t)\), \(T^{eq}_{yy}(t) = - G^{eq}(1,t)\), 
\(T^{eq}_{zz}(t)= 4 [M_z(t)]^2 -  G^{eq}(1,t) G^{eq}(-1, t)\), 
where \(G^{eq}(R,t)\), for \(R = \pm 1\), are given by 
\begin{eqnarray}
G^{eq}(R, t)  =      \frac{\gamma}{\pi}\int_{0}^{\pi} d \phi \cos(\phi R) \sin \phi 
               \frac{\mbox{tanh}\left(\frac{1}{2} \beta \Lambda(h(t))\right)}{\Lambda(h(t))} \nonumber \\
 - \frac{1}{\pi} \int_{0}^{\pi} d \phi \cos(\phi R) (\cos\phi -h(t))
               \frac{\mbox{tanh}\left(\frac{1}{2} \beta \Lambda(h(t))\right)}{\Lambda(h(t))}. \quad  
\end{eqnarray}

We can now calculate the equilibrium LN \(E_N^{eq}(t)\) of 
the two-site density matrix \(\rho^{eq}_{12}(t)\) of the equilibrium state \(\rho^{eq}(t)\) via the 
prescription in Eq. (\ref{eq_two_site_state}).
For the same values of the parameters as above,
so that 
\(\beta \approx 3.9\) via Eq. (\ref{energy}), 
the equilibrium LN of the two-site equilibrium state has an initial value of 
\(\approx 0.132635\). Then \(E_N^{eq}(t)\) jumps \emph{up} to \(\approx 0.157188\) for all \(t>0\). 
Note that the equilibrium magnetization \(M_z^{eq}(t)\) jumps \emph{down} in the same 
situation. So with more entanglement, we have more local disorder (cf. \cite{NK_babarey}).

Our interest is to compare this equilibrium two-site entanglement with the two-site entanglement 
of the evolved state. 
The nearest neighbor correlations of the evolved state are given by 
\cite{
McCoy_eka}
\(T_{xy} = S(1,t)/i\), \(T_{xx}(t) = -G(-1,t)\), \(T_{yy}(t) = -G(1,t)\), 
\(T_{zz}(t)= 4 [M_z(t)]^2 -  G(1,t) G(-1, t) + S(1,t)S(-1,t)\), 
where \(G(R,t)\) and \(S(R,t)\), for \(R = \pm 1\), are given by 
\begin{eqnarray}
\label{G_R}
 G(R, t)&=&  \frac{\gamma}{\pi}\int_{0}^{\pi} d \phi \sin(\phi R) \sin\phi 
               \frac{\mbox{tanh}\left(\frac{1}{2} \beta \Lambda(a)\right)}{\Lambda(a)\Lambda^2(b)} \nonumber \\
  \times  \big[\gamma^2 \sin^2 \phi &+& (\cos\phi -a) (\cos\phi -b) \nonumber \\
  && \quad \quad \quad \quad  +  (a-b) (\cos \phi-b) \cos(2\Lambda(b)t) \big] \nonumber \\ 
&-& \frac{1}{\pi} \int_{0}^{\pi} d \phi \cos(\phi R) 
               \frac{\mbox{tanh}\left(\frac{1}{2} \beta \Lambda(a)\right)}{\Lambda(a)\Lambda^2(b)} \nonumber \\
  \times  \big[\{\gamma^2 \sin^2 \phi &+& (\cos\phi -a) (\cos\phi -b)\} (\cos\phi -b)   \nonumber  \\
  && \quad \quad   -  (a-b) \gamma^2 \sin^2 \phi \cos(2\Lambda(b)t) \big], \\ 
S(R,t) &=&  \frac{\gamma (a-b)i}{\pi}\int_{0}^{\pi} d \phi \sin(\phi R) \sin\phi 
               \frac{\sin\left(2t\Lambda(b)\right)}{\Lambda(a)\Lambda(b)}. \nonumber \\
\end{eqnarray}

The  LN \(E_N(t)\) of the two-site density matrix \(\rho_{12}(t)\) of the 
evolved state \(\rho(t)\) can now be calculated by using Eq. 
 (\ref{evol_two_site_state}).
 The behavior of
LN of the evolved state 
for the same parameters as above,
 is shown in Fig. 
\ref{M_ent_XY}.
With time, 
the entanglement 
converges to a fixed value.
As \(\gamma\) approaches \(1\), the time taken for convergence is longer, until
for \(\gamma = 1\), 
the revivals occur for all times. 
Henceforth, we consider \(\gamma \ne 1\).  

As we see,  the two-site entanglement of the state obtained from  the time evolution 
(from an initial equilibrium state) is not approaching to the two-site entanglement 
of the equilibrium state for large times. 
This suggests that, just like the 
(single-site) magnetization,  the two-site entanglement of an infinite 
one-dimensional spin system in the XY model
is  \emph{nonergodic}.  

We will now compare these two nonergodic quantities of the XY chain. 
We will show that their temporal behaviors 
are in a sense complementary. 
First note that the magnetization \(M_z(t)\) of the evolved state 
is a damped oscillatory function. 
For our choice of parameters, the damping decreases the amplitude of the oscillation, but 
the mean value of the  oscillation is more or less fixed (see Fig. \ref{M_ent_XY}). 
Therefore, long time average of 
\(M_z(t)\) can essentially be considered to be a constant at \(\approx 0.02\),
after starting with an initial \emph{higher} value of \(\approx 0.148328\). 
Similar feature was seen before for \(M_z^{eq}(t)\).

The opposite is true when we compare the  two-site entanglements of the equilibrium state and 
the evolved states. The two-site LN \(E_N(t)\) of the evolved state converges to 
\(\approx 0.18\), after starting from an initial \emph{lower} value of \(\approx 0.132635\) 
(see Fig. 
\ref{M_ent_XY}). 
Again a similar behavior was seen in the equilibrium state.
It is in this sense that we say that the two-site entanglement and (single-site) 
magnetization has a complementary behavior in the infinite spin chain in the XY model.


The magnetization and nearest neighbor entanglement in the evolved state are 
plotted  
in Fig. \ref{M_ent_XY}. 
We note here that the next-nearest neighbor correlation functions are also known 
\cite{
McCoy_eka}.
Using them, we have calculated the LN for the next-nearest neighbor of 
the evolved state. 
It turns out that it is smaller than the nearest neighbor LN. For example,
for the parameters as above,
the next-nearest neighbor LN 
 is vanishing already at \(t = 0.8\) (and never going over \(\approx 0.0071\)), 
while the nearest neighbor LN 
is about \(0.18\), on average, for large times.

\begin{figure}[tbp]
\begin{center}
\epsfig{figure= 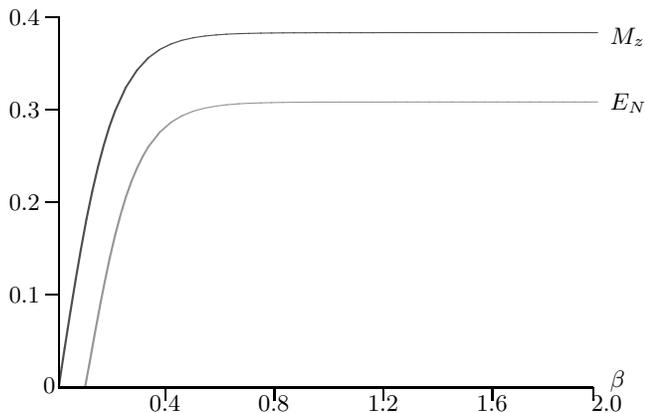,height=.2\textheight,width=0.4\textwidth}
\put(5,105){\(E_N\)}
\put(5,130){\(M_z\)}
\put(5,0){\(\beta\)}
\put(-203.5,0){\line(1,0){202}}
\put(-203.5,0){\line(0,1){140}}
\put(-209.5,-2){\(0\)}
\put(-169.3,-9){\(0.4\)}
\put(-163.3,0){\line(0,-1){5}}
\put(-122.1,0){\line(0,-1){5}}
\put(-128.1,-9){\(0.8\)}
\put(-80.9,0){\line(0,-1){5}}
\put(-86.9,-9){\(1.2\)}
\put(-39.7,0){\line(0,-1){5}}
\put(-45.7,-9){\(1.6\)}
\put(-1.5,0){\line(0,-1){5}}
\put(-2.3,-9){\(2.0\)}
\put(-203.5,35){\line(-1,0){5}}
\put(-222.5,32){\(0.1\)}
\put(-203.5,70){\line(-1,0){5}}
\put(-222.5,67){\(0.2\)}
\put(-203.5,105){\line(-1,0){5}}
\put(-222.5,102){\(0.3\)}
\put(-203.5,140){\line(-1,0){5}}
\put(-222.5,137){\(0.4\)}

\caption{The behavior in the evolved state, 
with temperature \(T\), of nearest neighbor entanglement  and magnetization 
of the infinite spin chain in XY model are compared for a fixed time \(t=1\) and 
for \(\gamma =0.5\), \(a = 10\), \(b=0\).
We actually plot the curves with respect to \(\beta\), where 
\(\beta = 1/kT\). 
}
\label{xy_ent_beta_mag}
 \end{center}
\end{figure}

Let us now consider the the behavior of the magnetization and nearest neighbor entanglement
of the evolved state, for different values of the temperature, but for a fixed time 
\(t = 1\). 
(See Fig. \ref{xy_ent_beta_mag}.) 
As expected, for high temperatures,
the (nearest neighbor) entanglement is vanishing.  Note, that both magnetization and 
nearest neighbor entanglement 
ultimately 
saturate, with decreasing \(T\). 

Summarizing, we have presented evidence 
that in the thermodynamic limit of an infinite one-dimensional chain of spin-1/2 particles 
described by the XY model Hamiltonian, the two-site entanglement is nonergodic. The two-site 
entanglement of the evolved state does not approach its equilibrium value. 
That is,  entanglement in such systems does not, by itself (i.e. without 
contact with external reservoirs),  relax to its equilibrium value, after a change of the external magnetic field.
This 
indicates that entanglement in such systems cannot be described by equilibrium statistical 
mechanics. 
 We also show that the 
entanglement has a complementary temporal behavior with respect to magnetization. 
We believe that such studies of the dynamics of entanglement in spin systems will help us to implement quantum 
information processing tasks in such systems 
(cf. \cite{Briegel_orey_orey}).

We acknowledge support of the Deutsche Forschungsgemeinschaft 
(SFB 407, SPP 1078), the 
Alexander von Humboldt  
Foundation, and
the EC Contract No. IST-2002-38877 QUPRODIS.


\begin{thebibliography}{10}

\bibitem{NC}M.A. Nielsen and I.L. Chuang, \emph{Quantum Computation and Quantum 
Information} (Cambridge University Press, Cambridge, 2000).













\bibitem{Osterloh}A. Osterloh
\emph{et al.},
Nature \textbf{416}, 608 (2002).


\bibitem{Nielsen}T. Osborne and M.A. Nielsen, Phys. Rev. A \textbf{66}, 032110 (2002). 

\bibitem{BEC}L.-M. Duan
\emph{et al.},
Phys. Rev. Lett. \textbf{85}, 3991 (2000); M.G.Moore and P. Meystre, Phys. Rev. Lett.
\textbf{85}, 5026 (2000). 

\bibitem{ent_length}  F. Verstraete, M. Popp, and J.I. Cirac,
Phys. Rev. Lett. \textbf{92}, 027901 (2004);   F. Verstraete, M.A. Martin-Delgado, and 
J.I. Cirac,
\emph{ibid.}, 087201 (2004).
      

\bibitem{Wootters} W.K. Wootters, Contemp. Math. \textbf{305}, 299 (2002); K.M. O'Connor and W.K. Wootters,
Phys. Rev. A, \textbf{63}, 052302 (2001);
G. Vidal
\emph{et al.},
Phys. Rev. Lett. \textbf{90}, 227902 (2003);
J. I. Latorre, E. Rico, and G. Vidal,  QIC \textbf{4}, 48 (2004), and references
therein.

\bibitem{numerical}  G. Vidal, quant-ph/0310089;
F. Verstraete, D. Porras, and J.I. Cirac, 
cond-mat/0404706;  S.R. Clark and D. Jaksch, cond-mat/0405580.

\bibitem{Briegel_orey_orey} R. Raussendorf and H. J. Briegel,
Phys. Rev. Lett. \textbf{86}, 5188 (2001). 


\bibitem{talk_dilo_Osterloh} L. Amico \emph{et al}., 
Phys. Rev. A \textbf{69},
022304 (2004); W. D{\" u}r
\emph{et al.},
quant-ph/0407075.

\bibitem{next-nearest} The  next-nearest (and further) neighbor  entanglement of the evolved state are smaller 
than the nearest neighbor one.


\bibitem{McCoy1}E. Barouch, B.M. McCoy and M. Dresden, Phys. Rev. A \textbf{2}, 1075 (1970).

\bibitem{LSM}E. Lieb, T. Schultz, and D. Mattis, Ann. Phys. (N.Y.) \textbf{16}, 407 (1961).




\bibitem{Gutzwiller}M.C. Gutzwiller, \emph{Chaos in Classical and Quantum Mechanics}
(Springer-Verlag, New York, 1990).


\bibitem{proposal_korechhey} 
J.J. Garcia-Ripoll and J.I. Cirac, Phil. Trans. R. Soc. Lond. A \textbf{361}, 1537 (2003); 
U. Dorner \emph{et al}., Phys. Rev. Lett. \textbf{91}, 073601 (2003);
L.-M. Duan, E. Demler, and M.D. Lukin, Phys. Rev. Lett. \textbf{91}, 090402 (2003).





\bibitem{McCoy_eka}
E. Barouch and B.M. McCoy, Phys. Rev. A \textbf{3}, 786 (1971);
\emph{ibid.}, 2137 (1971).










\bibitem{footnote123} Exploring the properties of entanglement in many-particle physical systems
is one of the  most important   research areas of quantum information processing and 
a variety of entanglement measures have been 
proposed (see e.g. 
\cite{MichalQIC}).

\bibitem{MichalQIC} 
C.H. Bennett
\emph{et al.},
Phys. Rev. A \textbf{54}, 3824 (1996); 
V. Vedral
\emph{et al.},
Phys. Rev. Lett \textbf{78}, 2275 (1997);
%
D.P. DiVincenzo \emph{et al}., 
  quant-ph/9803033;
 T. Laustsen, F. Verstraete, and S.J. van Enk,
QIC \textbf{3}, 64 (2003);
%
M.A. Nielsen,
Phys. Rev. Lett.
\textbf{83}, 436 (1999);
G. Vidal,
J. Mod. Opt. \textbf{47}, 355 (2000);
D. Jonathan and  M.B. Plenio, Phys. Rev. Lett. \textbf{83} 
 1455 (1999); 
 %
 M. Horodecki, A. Sen(De), and U. Sen, quant-ph/0403169;
 %
M. Horodecki, QIC \textbf{1}, 7 (2001).










\bibitem{VidalWerner}  G. Vidal and R.F. Werner, Phys. Rev. A, \textbf{65}, 032314 (2002).

\bibitem{Peres_Horodecki} A. Peres, Phys. Rev. Lett. \textbf{77}, 1413 (1996);
M. Horodecki, P. Horodecki, and R. Horodecki, 
Phys. Lett. A \textbf{223}, 1 (1996).



\bibitem{Horodecki_distillable} M. Horodecki, P. Horodecki, and R. Horodecki
Phys. Rev. Lett. \textbf{78}, 574 (1997).

\bibitem{hotei_hobey} This is a necessity, since the evolution of our spin chain starts with the initial 
equilibrium state, i.e. \(\rho(0) = \rho^{eq}(0)\).

\bibitem{Mazur} P. Mazur, Physica, \textbf{43}, 533 (1969).



\bibitem{two-site} Henceforth, unless it is explicitly stated otherwise, a two-site 
density matrix (or correlation or entanglement) will mean the corresponding ones for 
nearest neighbor sites of the infinite spin chain.

\bibitem{NK_babarey} M.A. Nielsen and J. Kempe, Phys. Rev. Lett. \textbf{86}, 5184 (2001).





\end{thebibliography}
\end{document}